# SQUARE GRID POINTS COVERAGED BY CONNECTED SOURCES WITH COVERAGE RADIUS OF ONE ON A TWO-DIMENSIONAL GRID

Pattama Longani

The Theory of Computation Group, Computer Science Department, Faculty of Science
Chiang Mai University, Chiang Mai, 50200, Thailand

## ABSTRACT

*We take some parts of a theoretical mobility model in a two-dimension grid proposed by Greenlaw and Kantabutra to be our model. The model has eight necessary factors that we commonly use in a mobile wireless network: sources or wireless signal providers, the directions that a source can move, users or mobile devices, the given directions which define a user's movement, the given directions which define a source's movement, source's velocity, source's coverage, and obstacles. However, we include only the sources, source's coverage, and the obstacles in our model. We define SQUARE GRID POINTS COVERAGE (SGPC) problem to minimize number of sources with coverage radius of one to cover a square grid point size of p with the restriction that all the sources must be communicable and proof that SGPC is in NP-complete class. We also give an APPROX-SQUARE-GRID-COVERAGE (ASGC) algorithm to compute the approximate solution of SGPC. ASGC uses the rule that any number can be obtained from the addition of 3, 4 and 5 and then combines 3-gadgets, 4-gadgets and 5-gadgets to specify the position of sources to cover a square grid point size of p. We find that the algorithm achieves an approximation ratio of $1 + \dfrac{2p-10}{p^2+2}$.*

*Moreover, we state about the extension usage of our algorithm and show some examples. We show that if we use ASPC on a square grid size of p and if sources can be moved, the area under the square grid can be covered in eight-time-steps movement. We also prove that if we extend our source coverage radius to 1.59, without any movement the area under the square gird will also be covered. Further studies are also discussed and a list of some tentative problems is given in the conclusion.*

## KEYWORDS

*Mobility Model, Wireless sensor Network, Coverage, Complexity, Wireless mobile communications.*

## 1. INTRODUCTION

Mobile wireless tools have become ubiquitous because they are of low cost, easy to use and able to fulfill many of the needs of human life. In [1] and [2], types of wireless networks, the usage fields and the survey of wireless technologies on wireless sensor network are specified. There are varieties of wireless sensor applications for cooperated collecting, monitoring and tracking information. There are a lot of models, experiments and algorithms to solve wireless sensor problems proposed. In [3], So and Ye apply Voronoi diagram to solve some of the coverage problems. In [4] Gabriele and Giamberardino propose mathematical dynamic sensor networks on discrete time. There are a lot of models and complex mathematical equations proposed. In [5] contexts about the coverage problem in wireless sensor network are stated. The authors explain three groups of the problems which are the area coverage, the point coverage and the path coverage. In [6] Jain and Sharma modify the swarm optimization algorithm which is an algorithm for placing nodes in wireless sensor network in a discrete space and propose *PSO* algorithm. They simulate the coverage area for some conditions to show that their algorithm is better than the swarm optimization algorithm. In [7], the authors define the coverage problems and experiment some cases of the problems by using techniques based on computational geometry, graph theory, Voronoi diagram and graph search algorithms. In [8], the authors explore the ways of shrinking some grids to increase the coverage area with the condition that sensors are allowed to move. Unlike most existing work, Greenlaw and Kantabutra model is the theoretical mobility model study wireless communication in a mobile environment from the perspective of complexity theory in a two-dimension grid, see [9], [10]. This model provides a framework factors that we commonly use in a mobile wireless network. Therefore, we choose this model.

The outline of this paper is as follows: basic definitions and notations are stated in Section 2. Our SQUARE GRID POINTS COVERAGE (*SGPC*) problem is also defined in this section. Section 3 states the proof to show that *SGPC* is in the *NP*-complete class. Section 4 presents APPROX-SQUARE-GRID-COVERAGE (*ASGC*), an approximation algorithm to compute an approximate solution to *SGPC*. Section 5 extends the approximation result in case that the sources can be moved and the sources radius are longer. Finally, conclusions and future studies are discussed in Section 6.

## 2. DEFINITIONS AND NOTATIONS

In this section, definitions and notations that we shall use are given: the grid, mobile wireless model, SQUARE GRID POINTS COVERAGE (*SGPC*) problem and source with coverage radius of one properties. Let $N = \{1, 2, 3, \ldots\}$ and let $S$ denote a set and $S^b$ the $b$-fold Cartesian product of $S$, that is, $S \times \cdots \times S$, where $S$ is repeated $b$ times. The basic definitions of the grid and our mobility model are given respectively as follows:

### 2.1 The Grid

Because we work on a two-dimensional grid, the position of a grid point will be on $(x, y)$, where $x, y \in N$. However, we number the points in a square grid from the left to the right and from the top to the bottom. Thus, point $(1, 1)$ is the top left point of a square grid.

**Definition 2.1.1 (Grid Points)** *A* grid point *is the point at the intersection between the equidistant interlocking perpendicular, vertical and horizontal axes.*

**Definition 2.1.2 (Square Grid)** *A* square grid *is the set of grid points on a two-dimensional grid in which the number of points in every x and y axes are equal.*

**Definition 2.1.3 (*Covered Grid Point, Covered Area*)** *A grid point is said to be* covered *by a source or sources if the point is within the coverage of the source or sources. We call the set of grid points covered by sources that are currently in range the* covered area*.*

**Definition 2.1.4 (Square Grid Size)** *Let $p \in N$. A square grid size of $p$ is a square grid which has the number of grid points on width * length equal to p * p points.*

A square grid has four equal sides and the points along each side are the boundaries for our interested area. We name the points on the boundary as follows:

**Definition 2.1.5 (Boundary Grid Point)** *The* boundary grid points *of a square grid are the set of the grid points which make the border of a square grid. The boundary grid points of a square grid size of p are $\{(1, i), (i, 1), (p, i), (i, p)\}$, $i \in N$ and $1 \le i \le p$.*

The boundary grid points are composed of *corner points* and *side points*.

**Definition 2.1.6 (Corner Point)** *A corner point is the boundary grid point at the corner of a square grid size of p * p. The corner points are the set of four points* {(1, 1), (1, p), (p, 1), (p, p)} *which we call the* top-left-corner-point, *the* top-right-corner-point, *the* bottom-left-corner-point, *and the* bottom-right-corner-point, *respectively.*

**Definition 2.1.7 (Side Point)** *A side point is any boundary grid point which is not a corner point. Because there are four sides in a square, we name them the* top-side-point, *the* bottom-side-point, *the* left-side-point, *and the* right-side-point.

## 2.2 Mobile Wireless Model

We shall focus on the mobility model $M = (S, D, U, L, R, V, C, O)$ of Greenlaw and Kantabutra. There are eight necessary factors that we commonly use in a mobile wireless network. The set $S$ is the set of *sources* or *wireless signal providers*. The set $D$ is the set of *directions* that a source can move in a two-dimensional grid in one *time step* of the movement $\tau$. The set $U$ is the set of *users* or *mobile devices*, a user cannot communicate directly to the other users, it must communicate via a source or a series of sources. The vector $L$ and $R$ are the given directions which define a *user's movement* and a *source's movement* respectively. Sources in this model can move with different velocities. The vector $V$ is a finite collection of numbers defines each source's *velocity*. Because Greenlaw and Kantabutra define every users' velocity to one, a user can move by one grid within a time, users do not have the set of velocities defined. Sources also have different coverages. The vector $C$ is a finite collection of numbers defines each source's *coverage*. The set $O$ is the set of *obstacles* which block a source signal, this model defines obstacles in a square shape.

Since sources must communicate with each other, Greenlaw and Kantabutra also define the communication protocol used in the model.

**Definition 2.2.1 (Coverage Representation)** *A coverage of radius c in a two-dimensional grid is represented by the set of grid points within the source coverage and on its boundary.*

**Definition 2.2.2 (Overlapping Coverage Area)** *Let s, s' be a coverage or an obstacle in a two-dimensional grid and s ∩ s' = z. We say that s overlaps s' if and only if |z| ≥ 2. z is called an overlapping coverage area.*

The manner in which they may communicate is specified as follows. Let $k > 2$ and $k \in \mathbb{N}$.

- At a given instance in time any pair of sources with overlapping-coverage areas may communicate with each other in full-duplex fashion as long as the intersection of their overlapping-coverage area is not completely contained inside obstacles. We say that these two sources are **currently in range**, or **communicable**. A series $s_1, s_2, ..., s_k$ of sources are said to be currently in range if $s_i$ and $s_{i+1}$ **are currently in range**, or **communicable**, for $1 \le i \le k - 1$.
- Two mobile devices cannot communicate directly with one another.
- A mobile device $D_1$ always communicates with another mobile device $D_2$ through a source or series of sources as defined next. The mobile devices $D_1$ at location $(x_1, y_1)$ and $D_2$ at location $(x_2, y_2)$ **communicate through a single source** $s$ located at $(x_3, y_3)$ if at a given instance in time the lines between points $(x_1, y_1)$ and $(x_3, y_3)$ and points $(x_2, y_2)$ and $(x_3, y_3)$ are within the area of coverage of $s$, and do not intersect with any obstacle from $O$. The mobile devices $D_1$ at location $(x_1, y_1)$ and $D_2$ at location $(x_2, y_2)$ **communicate through a series of sources** $s_1$ at location $(a_1, b_1)$, $s_2$ at location $(a_2, b_2)$,..., and $s_k$ at location $(a_k, b_k)$ that are currently in range if the line between points $(x_1, y_1)$ and $(a_1, b_1)$ is inside $s_1$'s coverage area and does not intersect any obstacle from $O$ and the line between points $(x_2, y_2)$ and $(a_k, b_k)$ is inside $s_k$'s coverage area and does not intersect any obstacle from $O$.

Because we are interested in only how to lay the minimum number of sources with a coverage radius of one in order to cover grid points sorted in a square shape, we reduce $M$ to be 3-tuples. So, our mobility model $M' = (S, C, O)$ is three tuples.

1. The set $S = \{s_1, s_2, ..., s_m\}$ is a finite collection of **sources**, where $m \in \mathbb{N}$. The value $m$ is the **number of sources**. Corresponding to each source $s_i$, for $1 \leq i \leq m$.

2. The vector $C = \{c_1, c_2, ..., c_m\}$ is a finite collection of lengths. The value $c_i$ is the corresponding radius of the circular coverage of source $s_i$. This vector is called the **coverages**. In this paper all $c_i = 1$, unless stated otherwise.

3. The set $O = \{(x_1, y_1, x_2, y_2) \mid x_1, y_1, x_2, y_2 \in N, x_2 > x_1 \text{ and } y_2 > y_1\}$ is a finite collection of rectangles in the plane. This set is called the **obstacles**. All coordinates $x_i$, $y_i \leq o_{max}$, where $o_{max}$ is a constant in N.

Because our sources also need to have communication with others, we use the same communication protocol which is used in the model $M$

## 2.3 Square Grid Points Coverage

We are interested in laying the minimum number of sources with coverage radius of one to cover a square grid with the restriction that all sources must be currently in range. The SQUARE GRID POINTS COVERAGE problem is defined as follows:

> SQUARE GRID POINTS COVERAGE (*SGPC*)
> INSTANCE: A mobility model $M'$, a square grid size of $p$, and a variable $k \in \mathbb{N}$.
> QUESTION: Can we lay at most $k$ sources to cover a square grid size of $p$ with the condition that all the sources must be communicable?

In *SGPC*, we are given a square grid size of $p$ with obstacles on some positions according to the given model $M'$. We are also given $k$ sources with coverage radius of one. We try to find whether we can lay this $k$ sources to cover all grid points with the condition that all sources must be communicable. Note: we focus on the sources that have the same coverage radius of one to cover a square and we can extend and vary the source coverage radius in the future work.

## 2.4 Sources with coverage radius of one

There are some properties of the sources with coverage radius of one. First of all, we note that a source radius of one centered on coordinate $(x, y)$ can cover exactly 5 grid points, namely, points on coordinates $(x+1, y)$, $(x-1, y)$, $(x, y+1)$, $(x, y-1)$, and $(x, y)$.

**Observation 2.4.1 (Maximum Points, Single Source)** A source with a coverage radius of one covers 5 grid points.

Suppose we have a source radius of one on coordinate $(x, y)$. Where can we place another source radius of one so that the two sources are currently in range (i.e., can communicate)? Obviously, the second source must be placed one unit apart in either $x$ or $y$ directions because the overlapping coverage area needs at least two grid points to communicate. Therefore, we can place another source radius of one on coordinate $(x + 1, y)$, $(x + 1, y + 1)$, $(x + 1, y - 1)$, $(x - 1, y)$, $(x - 1, y + 1)$, $(x - 1, y - 1)$, $(x, y + 1)$, or $(x, y - 1)$.

**Observation 2.4.2 (Pair Communication)** If two sources are not allowed to be centered on the same coordinate, source $s$ with coverage radius of one centered on $(x, y)$ can communicate with source $s'$ centered on coordinates $(x + 1, y)$, $(x + 1, y + 1)$, $(x + 1, y - 1)$, $(x - 1, y)$, $(x - 1, y + 1)$, $(x - 1, y - 1)$, $(x, y + 1)$, or $(x, y - 1)$.

Note: it follows from Observation 2.4.2 that any overlapping area between a pair of sources with coverage radius of one contains exactly two grid points.

**Observation 2.4.3 (Overlapping Coverage)** Any overlapping area between a pair of sources with coverage radius of one contains exactly two grid points.

From Observation 2.4.2, we also know that two sources are currently in range along an $x$-axis if there is at least one source centered on every $y$ coordinate according to Proposition 2.4.4.

**Proposition 2.4.4** Let $i, j, l \in \mathbb{N}$. Sources $s, s' \in S$ centered on $(x_i, y)$ and $(x_{i+j}, y)$, respectively. The two sources are currently in range along an $x$-axis if there is at least one source centered on every $(x_l, y)$, where $x_i < x_l < x_{i+j}$.
**Proof.** Suppose there is no source centered on $(x_l, y)$, $x_i < x_l < x_{i+j}$. As a result, the sources on $(x_{l-1}, y)$ and $(x_{l+1}, y)$ cannot communicate because the overlapping coverage area is less than two grid points. Therefore, along an $x$-axis there must be at least one source centered on every $x$-coordinate between $s$ and $s'$ to make all the sources between $s$ and $s'$ currently in range. □

In this section, we give the definitions and notations that we shall use in this paper. Next, we will show the hardness of the problem. The prove to show that the SQUARE GRID POINTS COVERAGE problem is in the $NP$-complete class is presented in the next section.

# 3 SGPC IS NP-COMPLETE

To solve $SGPC$, first, we try to find a polynomial time algorithm for solving the problem. However, we cannot find such an algorithm. In this section, we will show that $SGPC$ is in $NP$-complete class by making a reduction from an $NP$-complete problem called ONE-IN-THREE 3SAT in the case that no $c_i \in C, i \in \mathbb{N}$. contains a negated literal, see [11]. The definition of ONE-IN-THREE 3$SAT$ is defined as follows:

> ONE-IN-THREE 3SAT (*3SAT*)
> INSTANCE: Set $U$ of variables, collection $C$ of clause over $U$ such that each clause $c \in C$. has $/c/ = 3$.
> QUESTION: Is there a truth assignment for $U$ such that each clause in $C$ has exactly one true literal?

To prove that $SGPC$ is in $NP$-complete class, we shall show how to construct an $SGPC$ instance from a *3SAT* instance. We introduce six gadgets: Odd-gadget, Even-gadget, Square-gadget, N-odd-gadget, N-even-gadget and End-gadget as shown in Figure 1. The grey parts of the gadgets are the block of the obstacles.

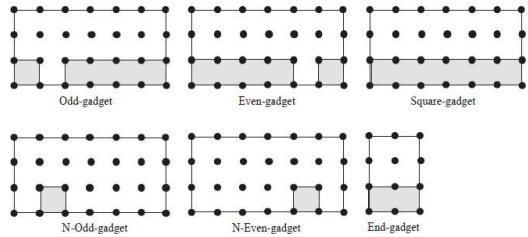

Figure 1.Six gadgets

We separate our building steps into three phases. In the first phase, we lay a row of Odd-gadgets or Even-gadgets equal to the number of variables, defined as /U/, along x-axes and switch laying the row of Odd-gadgets or Even-gadgets along y-axes equal to the number of clause, defined as |C/. Now, we have the switched rows of Odd-gadgets and Even-gadgets. We order each clause of 3*SAT* and the ascending order of the variables from the left to the right. We shall match each clause in 3SAT with each row of the obstacles in the gadgets, and match each variable in *3SAT* with each column of the gadgets. Next, if a variable is not in a clause, we will lay N-odd-gadget (N-even-gadget) over Odd-gadget (Even-gadget) along the column and the row of the gadgets which represent the variable and the clause. The N-odd-gadget (N-even-gadget) will block the hole in Odd-gadget (Even-gadget). We shall do this in every clause. From our building, there are holes only in the position of the gadget which represent the corresponding variable in a position, so there are exactly three holes along the row of obstacles. The second phase is to make our number of grid points to be the multiple of three for reducing the possible cases for which we shall clarify them later. From the first phase, there are $1 + 6|U|$ of grid points along each row. We add an End-gadget after every row of Odd-gadgets and Even-gadgets, then we have $3 + 6/U/ = 3(1 + 2/U/)$ of grid points on every row. For each column, we have $1 + 3/C/$ of grid points. We include two more grid points under each column, then we have $3 + 3|C| = 3(1 + |C/)$ grid points along each column. Now the number of grid points in our every row and column are the multiple of three.

Note: we do the second phase after the third phase, but we will keep in mind that the number of grid points along each row and column are the multiple of three. This means that at the end we will have two addition points under every column of grid points and we shall have an End-gadget after every row of the gadgets.

In the third phase, we shall make our area square. All gadgets, except for the End-gadget, cover 7 * 4 grid points. When we add a gadget to our area, we will extend 6 grid points in a row and extend 3 grid points in a column. Therefore, to make our area square, there are three cases possible. Let represent the number of the grid points along each row from the first and the second phase by h, and the number of the grid points along each column from the first and the second phase by v. If $h = v$, the grid points are already squared. If $h > v$, we shall add $\frac{h-v}{3} = 2|U|-|C|$ gadgets under each column, the switched of Odd-gadgets and Even-gadgets under the first column and the Square-gadgets under all other columns. Here, we will have a square grid size of $p = h = 3(1+2|U|)$. In case $v > h$, if 6 *mod* $(v-h) = 0$, we shall add $\frac{v-h}{6}$ of Square-gadgets after each row. Here, we will have a square grid size of $p = v = 3(1+|C|)$. If 6 *mod* $(v-h) \neq 0$, we shall add $\left\lfloor \frac{v-h}{6} \right\rfloor + 1$ of the Square-gadgets after each row and add one Odd-gadget or Even-gadget under the first column of the gadgets, continue the switching, and add one Square-gadget under all other columns. After three phases, we shall have a square grid size of $p = v + 3 = 3(|C| + 2)$.

This polynomial time construction is in $O(p^2)$. The left part of Figure 2 shows the reduction of a ONE-IN-THREE 3$SAT$ instance to an instance of $SGPC$. The 3$SAT$ instance used in the reduction is $(u_1 \vee u_2 \vee u_3)(u_1 \vee u_4 \vee u_5)(u_2 \vee u_3 \vee u_4)(u_1 \vee u_3 \vee u_4)(u_1 \vee u_2 \vee u_4)$.

$$(u_1 \vee u_2 \vee u_3)(u_1 \vee u_4 \vee u_5)(u_2 \vee u_3 \vee u_4)(u_1 \vee u_3 \vee u_4)(u_1 \vee u_2 \vee u_4)$$

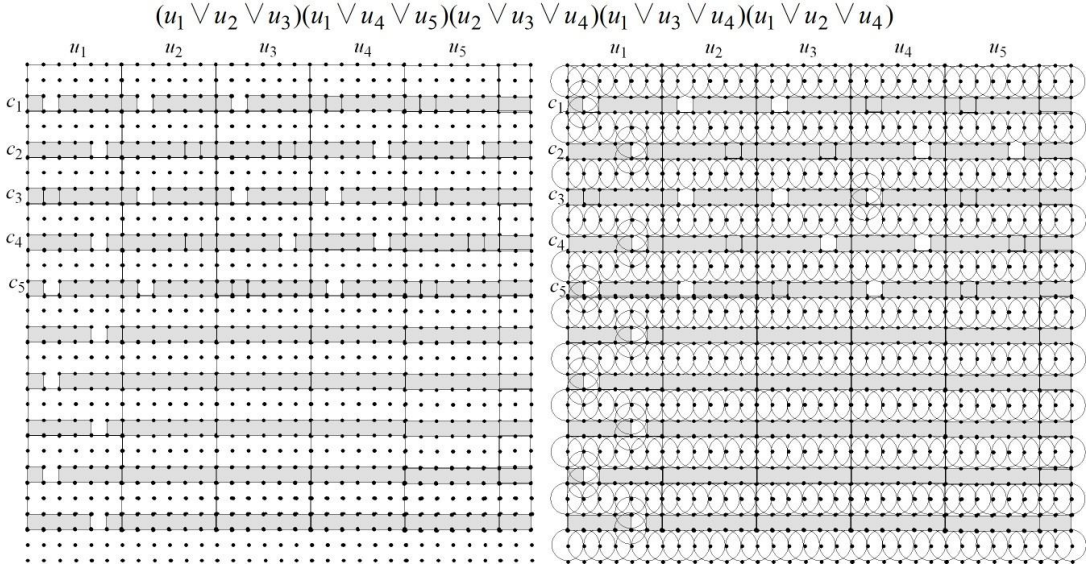

Figure 2. Reduction example

Next, the $NP$-complete proof is stated.

**Theorem 3.1** If $P \neq NP$, the Square Grid Points Coverage is $NP$-complete

**Proof.** We shall first show that $SGPC$ is in $NP$. Given a mobility model $M'$, a square grid size of $p$, a variable $k$, and a certificate which is a set of sources centered on some grid points in the square grid. We can verify the number of sources, the currently in range status among the sources, and the number of grid points that the sources cover in polynomial time of $O(p^2)$. Therefore, $SGPC$ is in $NP$.

Next, we shall show that is $SGPC$ is $NP$-hard. We claim that there is a truth assignment for $U$ that satisfies all the clauses in $C$ in exactly 1 in 3 manner if an only if $\frac{p^2}{3} + 2\left(\frac{p}{3} - 1\right)$ sources are currently in range and cover all grid points in a square grid size of $p$ with the condition that the first column will have the number of sources equal to $7\frac{p}{3} + 2b_i + 2\left(\left(\frac{p}{3} - 1\right) - |C|\right)$ or $7\frac{p}{3} + 2\left(\left(\frac{p}{3} - 1\right) - |C|\right)$ and the other columns will have the number of sources equal $7\frac{p}{3} + 2b_i$ or $7\frac{p}{3}$. The variable $b_i$, $b_i \in \mathbb{N}$, is the number of the clauses that a variable $u_i \in U$ appears.

→ Suppose there is a truth assignment for $U$ that satisfies all the clauses in $C$ in exactly 1 in 3 manner, therefore, exactly one variable in a clause is assigned true. In order to find the corresponding answer we lay rows of connected sources over and under every obstacle to cover all grid points. Because the square grid size is equal to $p$, there are $p$ sources in the rows, as stated in Proposition 2.4.4. Because there are $\dfrac{p}{3}$ rows, we shall use $\dfrac{p^2}{3}$ sources. However, these sources in the different rows cannot be communicable. From Proposition 2.4.4, we must have at least one source on every $y$-axis to make all the sources communicable, so we add two sources for connecting any pair of rows of connected sources. Because there are $\left(\dfrac{p}{3}-1\right)$ rows of the obstacles, we shall use $2\left(\dfrac{p}{3}-1\right)$ sources for the connection. Therefore, we must use $\dfrac{p^2}{3}+2\left(\dfrac{p}{3}-1\right)$ sources to cover the square grid size of $p$. For laying sources, we shall lay the two sources at a hole on a rows of obstacles. Because there are three holes in every row of the obstacles which represent the corresponding clause and three variables in 3$SAT$, only one hole will be used for the connection corresponding with only one truth assignment of a variable in a clause of 3$SAT$. If a variable is assigned true, we will put the two sources in every hole in the column that represent the variable, for the consistency assignment value to the variable in every clause. If a variable is assigned false, there are no two sources in any hole in the column that represent the variable. Let the number of the clause that a variable $u_i \in U$ appears in is equal to $b_i$, $b_i \in$ N. If 3$SAT$ answer is "yes", each column corresponding to a variable will has the number of sources equal to $7\dfrac{p}{3}+2b_i$ or $7\dfrac{p}{3}$ when the variable is assigned true or false, respectively.

However, the first column which has more holes than the number of clauses will have the number of sources equal to $7\dfrac{p}{3}+2b_i+2\left(\left(\dfrac{p}{3}-1\right)-|C|\right)$ or $7\dfrac{p}{3}+2\left(\left(\dfrac{p}{3}-1\right)-|C|\right)$ when the variable is assigned true or false, respectively. (If 3$SAT$ answer is "no", we cannot make all sources currently in range, or the number of sources in a column that represent variable $u_i$ is not exactly equal to $7\dfrac{p}{3}+2b_i$ or $7\dfrac{p}{3}$, or the number of sources in the first column is not equal to $7\dfrac{p}{3}+2b_i+2\left(\left(\dfrac{p}{3}-1\right)-|C|\right)$ or $7\dfrac{p}{3}+2\left(\left(\dfrac{p}{3}-1\right)-|C|\right)$, or we are using more than $\dfrac{p^2}{3}+2\left(\dfrac{p}{3}-1\right)$ sources to cover all grid points with the condition that all the sources are currently in range.)

←We have $\dfrac{p^2}{3}+2\left(\dfrac{p}{3}-1\right)$ sources to cover a square grid size of $p$. To cover all grid points, from Proposition 2.4.4, we will use $p$ sources at the minimum to cover. So, there are $2\left(\dfrac{p}{3}-1\right)$ sources left and a source in a row of connected sources are not currently in range with the sources in the other rows. Because there are $\dfrac{p}{3}-1$ rows of obstacles, we will use exactly two sources to connect two rows of connected sources at the maximum. Because there are three holes in a row of obstacles, we can put two sources in any hole to connect two rows of connected sources. However, if we do not lay sources in every hole in the same column and do not lay no source in every hole in the same column, the number of source will not exactly equal to $7\dfrac{p}{3}+2b_i+2\left(\left(\dfrac{p}{3}-1\right)-|C|\right)$ or $7\dfrac{p}{3}+2\left(\left(\dfrac{p}{3}-1\right)-|C|\right)$ for the first column and will not exactly equal $7\dfrac{p}{3}+2b_i$ or $7\dfrac{p}{3}$ for the other columns. Thus, we must lay two sources in every hole in the same column or not lay any source in every hole in the same column. We will assign the corresponding variable in a column to true if we lay two sources in every hole in the column, and will assign the corresponding variable in a column to false if we lay two sources in every hole in the column that we do not lay any source in every hole of the column. Because we will lay two sources in one of three holes in a row of obstacles and we will lay two sources in every hole in the same column or do not lay any source in every hole in the same column, therefore these correspond to the truth assignment of $3SAT$ instance in exactly 1 in 3 manner. ☐

On the right part of Figure 2, an example of wrong answer of $SGPC$ is shown. Although the answer use $\frac{33}{3} * 33 + 2\left(\frac{33}{3} - 1\right)$ sources with the condition that all sources are currently in range, the number of sources in the column which represented the variable $u_4$ is not equal to $7 * \frac{33}{3}$ or $7 * \frac{33}{3} + 2 * 4$, $b_4 = 4$, sources.

In this section, we show that $SGPC$ is $NP$-complete. Next, we shall show a way to manage $SGPC$.

## 4. BOUND AND APPROXIMATION ALGORITHM FOR $SGPC$

The approximation algorithm for $SGPC$ is presented. The algorithm use the rule that any number can be reached through the addition of 3, 4 and 5 and then combine 3-gadgets, 4-gadgets and 5-gadgets to specify the position of sources to cover a square grid point size of p and return the number of the sources used. In section 4.1, the lower bound number of communicable sources to cover a square grid size of $p$ is shown and in section 4.2 APPROX-SQUARE-GRID-POINTS-COVERAGE is presented.

### 4.1 The lower bound number of sources with coverage radius of one covering a square grid

When we lay a source on the boundary of a square grid, the source will cover some points outside our interested area. From the following propositions, the number of sources in our three gadgets are optimal.

**Proposition 4.1.1** *When p >1, A source centered on a corner point of a square grid covers* 3 *grid points in the square grid.*
**Proof:** There are four corner points in a square grid. Suppose $s$ is an arbitrary source centered on $(x, y)$. If $s$ is centered on the top-left-corner-point, $s$ will cover two points outside the square grid area, i.e. on the positions $(x, y - 1)$ and $(x - 1, y)$. When $s$ is centered on the top-right-corner-point, the bottom-left-corner-point and the bottom-right-corner-point, $s$ will also cover two points outside the square grid area. Therefore, a source centered on a corner point of a square grid will cover 3 grid points of the square grid when $p >1$. ☐

**Proposition 4.1.2** *A source centered on a side point of a square grid covers* 4 *grid points in the square grid when p > 2.*
**Proof.** There are four side points. Suppose $s$ is an arbitrary source centered on $(x, y)$. If $s$ is centered on one of the top side points, $s$ will cover one point outside the square grid area on the position $(x, y - 1)$. When $s$ is centered on outside the bottom side points, one of the left side points and one of the right side points, $s$ will also cover one point outside the square grid area. Therefore, a source centered on a side point of a square grid covers 4 grid points in the square grid when $p > 2$. ☐

**Proposition 4.1.3** *If we lay a source to cover a corner point of a square grid, the source will cover at most one point on a square grid size of one, cover at most three points on a square grid size of two and cover at most four points on a square grid size of $p \geq 3$.*

**Proof.** If a corner point is on $(x, y)$, a source with coverage radius of one must be centered not further than $x \pm 1$ and $y \pm 1$ coordinate. By considering laying a source on five possible coordinates to cover a corner point which are $(x + 1, y)$, $(x - 1, y)$, $(x, y)$, $(x, y + 1)$, $(x, y - 1)$, it can be seen that the proposition holds.

**Proposition 4.1.4** *At least one source is needed to cover a square grid point size of one. (a grid point)*

**Proof.** A grid point must be covered by a source. □

**Proposition 4.1.5** *At least two sources are needed to cover a square grid size of two.*

**Proof.** Let $x, x', y, y' \in \mathbb{N}$. From Proposition 3.1, a source with coverage radius of one centered on $(x, y)$ position will cover five grid points which are $(x + 1, y)$, $(x - 1, y)$, $(x, y)$, $(x, y + 1)$, and $(x, y - 1)$. There are four grid points in a square grid size of two which are $(x', y')$, $(x' + 1, y')$, $(x', y' + 1)$, $(x' + 1, y' + 1)$. It is impossible to map all points of a source to the points on a square grid size of two. Therefore, we must use at least two sources with coverage radius of one to cover a square grid size of two. □

**Proposition 4.1.6** *At least three communicable sources are needed to cover a square grid size of three.*

**Proof.** There are nine grid points in a square grid size of three. Suppose we can use less than three communicable sources. However, from [12] it is stated that two sources that are currently in range can cover $(3*2) + 2 = 8$ grid points at the maximum, so any pair of sources that are currently in range cannot cover a square grid size of three. Thus, we must use at least three communicable sources. □

**Theorem 4.1.7** At least $\left\lceil \dfrac{p^2 + 2}{3} \right\rceil$ communicable sources are needed to cover a square grid size of $p$, $p \geq 4$.

**Proof.** Suppose we can use $\left\lceil \dfrac{p^2 + 2}{3} \right\rceil - k$, $k \geq 1 \in \mathbb{N}$, communicable sources with coverage radius of one to cover a square grid size of $p$. From [12] it is stated that $m$ sources can cover $3m+2$ grid points at the maximum, $\left\lceil \dfrac{p^2 + 2}{3} \right\rceil - k$ sources that are currently in range will cover $3*\left(\left\lceil \dfrac{p^2 + 2}{3} \right\rceil - k\right) + 2 = 3*\left\lceil \dfrac{p^2 + 2}{3} \right\rceil - 3k + 2$ grid points which is the maximum. However, from the Proposition 4.3, to cover a corner point, a source will cover at most four points in a square grid when $p \geq 3$. Because there are four corners in a square grid, there are at least four uninterested points covered by the communicable sources. Thus, $\left\lceil \dfrac{p^2 + 2}{3} \right\rceil - k$ communicable sources can cover only $3*\left\lceil \dfrac{p^2 + 2}{3} \right\rceil - 3k + 2 - 4 = 3*\left\lceil \dfrac{p^2 + 2}{3} \right\rceil - 3k - 2$ grid points on a square grid. Suppose $3*\left\lceil \dfrac{p^2 + 2}{3} \right\rceil - 3k - 2 \geq p^2$ where $p$ is the size of a square grid, we have $k \leq \left\lceil \dfrac{p^2 + 2}{3} \right\rceil - \dfrac{p^2 + 2}{3}$. We know that $\left\lceil \dfrac{p^2+2}{3} \right\rceil < \dfrac{p^2+2}{3}+1$. Therefore, $k < 1$ which contradicts the fact that $k \geq 1$. Thus, we must use at least $\left\lceil \dfrac{p^2+2}{3} \right\rceil$ connected sources with coverage radius of one to cover a square grid size of $p \geq 4$.  □

## 4.2 APPROX-SQUARE-GRID-COVERAGE (*ASGC*)

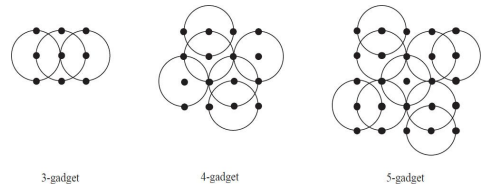

Figure 3. 3-gadgets, 4-gadgets and 5-gadgets

Because *SGPC* is in the *NP*-complete class, we cannot find a polynomial time algorithm to solve the problem. Therefore, in this section, we propose an algorithm called Approx-SQUARE-GRID-COVERAGE (*ASGC*) to lay sources on a square grid size of $p$ when $p > 5$. Because any number can be obtained from the addition of 3, 4 and 5, we introduce 3-gadgets, 4-gadgets and 5-gadgets which cover a square grid size of 3, 4 and 5 respectively, as shown in Figure 3. From Proposition 4.1.6, the number of sources which coverage radius of one on the 3-gadget is optimal, i.e. three sources. From Theorem 4.1.7, to cover a square grid size of 4 and 5 we must use at least $\dfrac{4^2+2}{3} = 6$ and $\dfrac{5^2+2}{3} = 9$ sources, respectively. Therefore, the six sources in a 4-gadget and the nine sources in a 5-gadget are optimal. We shall combine these three types of gadgets to cover any square grid size of $p > 5$. Because any number larger than five can be written through the addition of 3, 4 and 5, the functions in *ASGC* is classified into three cases according to three possible fractions of dividing $p$ by 3.

---

APPROX-SQUARE-GRID-POINTS-COVERAGE($p$)
1: **if** 3 *mod* $p$ equal to 0 **then**
2:      FRACTIONZERO($p$)
3: **else if** 3 *mod* $p$ equal to 1 **then**
4:      FRACTIONONE($p$)
5: **else if** 3 *mod* $p$ equal to 2 **then**
6:      FRACTIONTWO($p$)
7: **end if**

---

Table 1. APPROX-SQUARE-GRID-POINTS-COVERAGE Pseudocode

We shall call the series of grid points on the same $y$ ($x$) axis "*a row (column) of grid points,*" and call the connected 3-gadgets where all sources are centered on the same $y$ ($x$) axis "*a row (column) of connected sources.*" In case 3mod $p = 0$, FRACTIONZERO shall be used. In this case, $p$ can be written as the sum of 3's. Therefore, we use only 3-gadgets to cover our square grid points by laying 3-gadgets along each three rows of grid points. Let the variables $r$ and $c$ represent the number of the row of grid points and the column of grid points respectively. FRACTIONZERO will lay the top-left-corner-point of the 3-gadgets on every other three rows of grid points which have $r = 1, 4, 7, \ldots , p - 2$, and lay the top-left-corner-point on every other three columns which have $c = 1, 4, 7, \ldots , p - 2$, respectively. Therefore, there are $\dfrac{p}{3}$ rows and $\dfrac{p}{3}$ columns to lay the 3-gadgets and we use $3 * \dfrac{p}{3} * \dfrac{p}{3} = \dfrac{p^2}{3}$ sources to cover the square grid points size of $p$. Because the sources in the different rows of connected sources are uncommunicable, we add two more sources to connect any two rows of connected sources. We center the two sources on $(2, r)$ and $(2, r + 1)$, respectively, and use $2 * \left( \dfrac{p}{3} - 1 \right) = 2 * \dfrac{p}{3} - 2$ sources. However, we can delete one source in each row of connected sources that is in between the addition source while all the sources are still communicable and we shall delete the same kind of sources in every row of connected sources except the first and the last rows. Therefore, we can delete $\dfrac{p}{3} - 2$ sources.

As a result, we uses $\left( \dfrac{p^2}{3} \right) + 2 * \left( \dfrac{p}{3} \right) - 2 - \left( \dfrac{p}{3} - 2 \right) = \dfrac{p(p+1)}{3}$ sources to cover a square grid point size of $p$ and FRACTIONZERO takes $O(p^2)$ which is in time polynomial. The pseudocode of FRACTIONZERO is shown in Table 2.

---

FRACTIONZERO($p$)
1: //lay 3-gadget along each three rows of grid points:
2: $r, c \leftarrow 1$
3: **for** $r \leftarrow 1$ to $p - 2$ **do**
4:     **for** $c \leftarrow 1$ to $p - 2$ **do**
5:         lay 3-gadget on $(c, r)$
6:         $c = c + 3,$
7:     **end for**
8:     $r = r + 3$
9: **end for**
10: //add two sources to connect each two rows of connected sources and delete some    redundant sources
11: **for** $r \leftarrow 2$ to $p - 7$ **do**
12:     lay sources on $(2, r)$ and $(2, r + 1)$
13:     delete source on $(2, r + 2)$
14:     $r = r + 3$
15: **end for**
16: lay sources on $(2, r)$ and $(2, r + 1)$

---

Table 2. Pseudocode of FRACTIONZERO

The position of the 3-gadget and some example of laying the gadgets on some square grid are shown in Figure 4. Figure 4-A represents the 3-gadget. Figure 4-B shows a box represents a 3-gadget. The end of the two arrows pointing out of the box represent the position of the points the 3-gadget cover that are outside the boundary of the square grid size of 3 that a 3-gadget cover. Figure 4-C shows the position of laying 3-gadgets that cover a square grid size of 9. Figure 4-D and Figure 4-E show the sources position obtained from FRACTIONZERO to cover square grid size of 6 and 9 respectively. Two additional sources which connect any two row of connected sources are shown by the thick circles. The shaded source is the source that will be deleted, we will delete the same kind of the shaded sources in every row of connected sources except the first and the last rows.

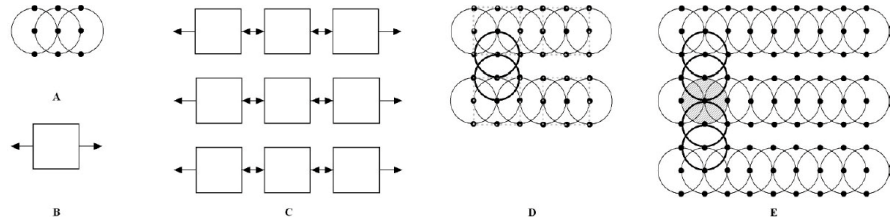

Figure 4. The example of laying 3-gadgets to cover a square grid size of 6 and 9.

In case $3 \bmod p = 1$, FRACTIONONE shall be used. This case $p$ can be written as an addition of some 3s and a 4, so we use 3-gadgets and a 4-gadget. First, we lay 4-gadgets along a diagonal line in a square grid. Let the variable $c$ represent the column number. FRACTIONONE will lay the top-left-corner-point of 4-gadgets along a diagonal line in a square grid on the position $(c, c)$ where $c = (1, 4, 7, ..., p - 3)$. We also vertically mirror every other 4-gadgets to make all the 4-gadgets communicable. In this part, we use $\frac{p-1}{3}$ of 4-gadgets. Because each 4-gadget has 6 sources, we use $\left(\frac{p-1}{3}\right)*6 = 2(p-1)$ sources. The positions of laying 4-gadgets will separate uncovered grid points into the left side and the right side of the communicable 4-gadgets. On the left side of the 4-gadgets, we start laying the top-left-corner-point of one 3-gadget on row number equal to 5, we increase laying one gadget in every row until the row number is equal to $p - 6$ and have $\frac{p-4}{3}$ of 3-gadgets laid in this row. Therefore, we use $\sum_{n=1}^{\frac{p-4}{3}} n$ of 3-gadgets or use $3*\sum_{n=1}^{\frac{p-4}{3}} n$ sources. On the right side of the 4-gadgets, we start laying $\frac{p-4}{3}$ of 3-gadgets on column number equal to 5, we increase column number by three and decrease one gadget in every row until there is only one 3-gadgets in the last row. Therefore, we use $\sum_{n=1}^{\frac{p-4}{3}} n$ of 3-gadgets or use $3*\sum_{n=1}^{\frac{p-4}{3}} n$ sources. The number of sources in the 3-gadgets laying on both side of the 4-gadgets are equal. As a result, we use $2(p-1)+2*3*\sum_{n=1}^{\frac{p-4}{3}} n = \frac{(p-1)(p+2)}{3}$ sources to cover a square grid points size of $p$ and FRACTIONONE takes $O(p^2)$ which is in time polynomial. The pseudocode of FRACTIONONE is shown in Table 3.

```
FRACTIONONE(p)
1: // lay 4-gadget along diagonal line:
2: c ← 1
3: for c ← 1 to p − 3 do
4:      lay 4-gadget on (c, c)
5:      FLIP(4-gadget)
6:      c = c + 3
7: end for
8: // lay 3-gadget along each three rows of grid points on the left of the 4-gadgets:
9: c_max ← 1
10: while c_max ≤ p − 6 do
11:      for c ← 1 to c_max do
12:          lay 3-gadget on (c, c_max + 4)
13:          c = c + 3
14:      end for
```

```
15:        c_max = c_max + 3
16: end while
17: // lay 3-gadget along each three rows of grid points on the right of the 4-gadgets
18: c_min ← 5
19: while c_min ≥ p − 2 do
20:        for c ← c_min to p − 2 do
21:            lay 3-gadget on (c, c_min − 4)
22:                c = c + 3
23:        end for
24:        c_min = c_min + 3
25: end while
```

Table 3. Pseudocode of FRACTIONONE

The position of the 3-gadgets and 4-gadgets and an example of laying the gadgets on a square grid size of 7 are shown in Figure 5. Figure 5-A represents a 4-gadget. Figure 5-B shows a box represents a 4-gadget. The end of the arrows pointing out of the box represent the position of the points the 4-gadget cover that are outside of the boundary of the square grid size of 4 that a 4-gadget cover. Figure 5-C shows the position of laying 3-gadgets and 4-gadgets to cover a square grid size of 10. Figure 5-D shows the sources positions obtained from FRACTIONONE to cover square grid size of 7.

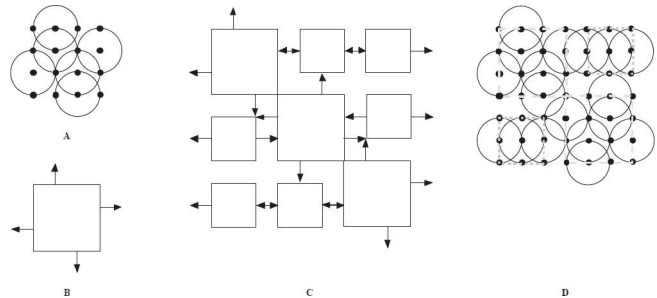

Figure 5. Examples of laying 3-gadgets and 4-gadgets to cover square grids size of 10 and 7

In case $3 \bmod p = 2$, FRACTIONTWO shall be used. This case $p$ can be written as an addition of some 3s and a 5, so we use 3-gadgets and a 5-gadget. First, we lays 5-gadgets along a diagonal line in a square grid. Let the variable $c$ represent the column number. FRACTIONTWO lay the top-left-corner-point of 5-gadgets along a diagonal line in a square grid on the position $(c, c)$ where $c$ = (1, 4, 7, ... , $p - 4$). In this part, we use $\frac{p-2}{3}$ of 5-gadgets. Since each 5-gadget has 9 sources, we use $\frac{p-2}{3} * 9 = 3(p-2)$ sources. The positions of laying 5-gadgets will separate uncovered grid points into the left side and the right side of the communicable 5-gadgets. On the left side of the 5-gadgets, we start laying the top-left-corner-point of one 3-gadget on row number equal to 6, we increase laying one gadget in every row until the row number is equal to $p - 7$ and have $\frac{p-5}{3}$ of 3-gadgets laid in this row. Therefore, we use $\sum_{n=1}^{\frac{p-5}{3}} n$ of 3-gadgets or use $3 * \sum_{n=1}^{\frac{p-5}{3}} n$ sources. On the right side of the 5-gadgets, we start laying $\dfrac{p-5}{3}$ of 3-gadgets on column number equal to 6, we increase column number by three and decrease one gadget in every row until there is only one 3-gadgets in the last row. Therefore, we use $\sum_{n=1}^{\frac{p-5}{3}} n$ of 3-gadgets or use $3*\sum_{n=1}^{\frac{p-5}{3}} n$ sources. The number of sources in the 3-gadgets laying on both sides of the 5-gadgets are equal. As a result, we use $2(p-1)+2*3*\sum_{n=1}^{\frac{p-5}{3}} n = \dfrac{(p-2)(p+4)}{3}$ sources to cover a square grid points size of $p$ and FRACTIONTWO will take $O(p^2)$ which is in time polynomial. The pseudocode of FRACTIONONE is shown in Table 4.

```
FRACTIONTWO(p)
1: // lay 5-gadget along diagonal line:
2: for c ← 1 to p − 4 do
3:      lay 5-gadget on (c, c)
4:      c = c + 3
5: end for
6: // lay 3-gadget along each three rows on the left of the 5-gadgets
7: c_max ← 1
8: while c_max ≤ p − 7 do
9:      for c ← 1 to c_max do
10:         lay 3-gadget on (c, c_max + 5)
11:         c = c + 3
12:     end for
13:     c_max = c_max + 3
14: end while
15: // lay 3-gadget along each three rows on the right of the 5-gadgets
16: c_min ← 6
17: while c_min ≥ p − 2 do
18:     for c ← c_min to p − 2 do
19:         lay 3-gadget on (c, c_min − 5)
20:         c = c + 3
21:     end for
22:     c_min = c_min + 3
23: end while
```

Table 4. The Pseudocode of FRACTIONTWO

The position of the 3-gadgets and 5-gadgets and some examples of laying the gadgets on some square grids are shown in Figure 6. Figure 6-A represents a 5-gadget. Figure 6-B shows a box represents a 5-gadget. The end of the arrows pointing out of the box represent the position of the points the 5-gadget cover that are outside of the boundary of the square grid size of 5 that a 5-gadget cover. Figure 6-C shows the position of laying 3-gadgets and 5-gadgets to cover a square grid size of 11. Figure 6-D shows the source positions obtained from FRACTIONTWO to cover square grid size of 8.

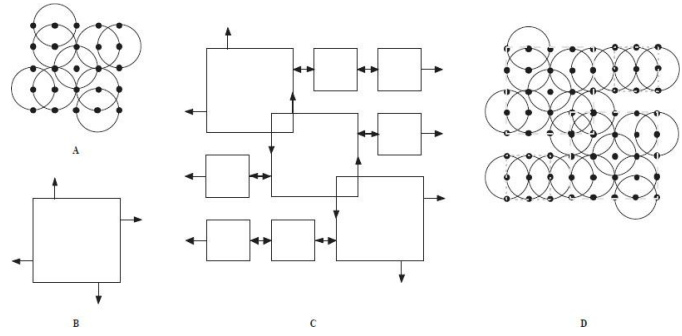

Figure 6. Examples of laying 3-gadgets and 5-gadgets to cover square grids size of 11 and 8

Next, we will show the approximation ratio of the algorithm.

**Theorem 4.2.1** APPROX-SQUARE-GRID-POINTS-COVERAGE is $\left(1 + \dfrac{2p - 10}{p^2 + 2}\right)$-*approximation algorithm for the* SQUARE GRID POINTS COVERAGE *problem when p>5*

**Proof.** We have already shown that *AGPC* runs in polynomial time. Let $|S^*|$ denote the minimum number of sources to cover a square grid size of $p$. Let $|A|$ denote the number of sources used to cover a square grid size of $p$ obtained from *AGPC*. From considering all three cases in *AGPC*, we can conclude that $\dfrac{p(p+1)}{3} \leq |A| \leq \dfrac{(p-2)(p+4)}{3}$. However, because the optimal number of sources must be less than or equal to the minimum number obtained from *AGPC*, so $|S^*| \leq \dfrac{p(p+1)}{3}$. It follows that $\dfrac{p^2+2}{3} \leq |S^*| \leq \dfrac{p(p+1)}{3} \leq |A| \leq \dfrac{(p-2)(p+4)}{3}$. Then we have $\dfrac{|A|}{|S^*|} \leq \dfrac{(p-2)(p+4)}{3|S^*|}$. From Theorem 4.1.7, we know that $\dfrac{p^2+2}{3} \leq |S^*|$. Therefore, the approximation ratio is equal to $\dfrac{|A|}{|S^*|} \leq \dfrac{(p-2)(p+4)}{p^2+2}$ or $\dfrac{|A|}{|S^*|} \leq 1 + \dfrac{2p-10}{p^2+2}$. $\qquad\square$

So far, we have proved that the *SGCP* is in *NP*-complete class and give an approximate algorithm with the approximation ration equal to $1 + \dfrac{2p-10}{p^2+2}$. Next, we shall show some works that are extended from the result of the approximation algorithm that we have proposed in this section.

## 5. EXTENSIONS

We have proposed *SGPC* and show that the problem is in the *NP*-complete class. We also propose *ASGC* algorithm to give an approximation value of *SGPC*. In this section, we will show that if the sources can be moved by one grid within a time, the sources will cover a square area size of $p$ within 8 time steps of movement. Moreover, we shall show that if we extend the radius of our sources coverage to 1.59, all the sources will cover the area under any square grid size of $p$.

## 5.1 Eight-Time Movement

The mobility model of Greenlaw and Kantabutra is a theoretical mobility model. They defines the possible directions that a source's movement in a two-dimensional grid in a time step: no movement, east, west, south, and north. We shall show that if a source move in *eight-time-steps fashion*: move to the north → south → south → north → east → west → west → east, respectively, the source will cover a square area size of *p* within 8 time steps of movement.

**Definition 5.1.1 (Square Area Size of *p*)** *A square area size of p is the area under a square grid size of p+1.*

Suppose source *A* with coverage radius of one is in a square grid size of three as shown in Figure 7-A. After *A* move to the north with one step, move to the south with two steps, move one step to the north and one step to the west, move two steps to the east, the area that sources cover and do not cover in each steps are shown in Figure 7-B, 7-C, 7-D, and 7-E, respectively.

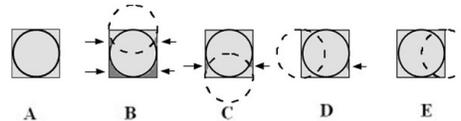

Figure 7.eight-time-steps fashion

**Observation 5.1.2** *If we move a source with coverage radius of one centered on (x,y) over the square grid size of three which has its top-left-corner-point on (x-1,y) in eight-time-steps fashion, the source will cover the square area under the square grid size of three.*

Suppose we have three sources and a square area as shown in Figure 8. If the sources move three steps: move one step to the north and move two steps to the south, shown in Figure 8-B and Figure 8-C, respectively, the square area size of one will be covered. Note: these three steps movement is also be a part of the eight-time-steps fashion.

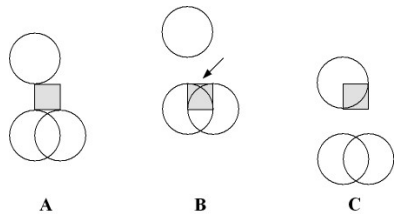

Figure 8. Covering a square area size of one

**Observation 5.1.3** *Suppose we have a square area size of one which have the top-left-corner-point on (x,y) and three sources centered on (x,y+1), (x,y-1), (x+1, y-1) respectively. If we move the three sources in eight-time-steps fashion, the square area size of one will be covered. It is also true if we rotate the three sources in the corresponding manner and move the sources in three steps in the corresponding manner.*

**Theorem 5.1.4** *If we move sources under ASPC according to eight-time-step fashion, the square area under a square grid size of p will be covered.*

**Proof:** Because the location of sources to cover a square grid is systematically specified, by consideration all the spaces possible according to Observation 5.1.2 and 5.1.3, all the square area under the square grid will be covered.  □

## 5.2 Sources with coverage radius of 1.59

In this section, we shall show that if we extend the radius of our sources coverage to 1.59, all the sources that we use according to *ASGC* will cover the area under any square grid size of *p*.

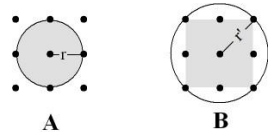

Figure 9. *A)* source with coverage radius of one. *B)* source with coverage radius of $\sqrt{2}$

Let $r \in \Re$. The grey part in Figure 9-B show that source with coverage radius of 1.414 can cover the square area size of two.

**Observation 5.2.1** Sources with coverage radius of $\sqrt{2}$ centered on a grid point will cover the square area size of two.

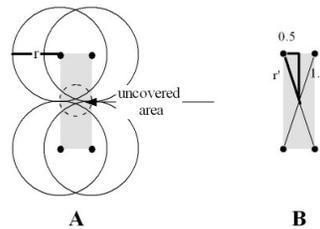

Figure 10. *A)* The area that is not covered by the four sources with r = 1.5

$$B) \ r' = \sqrt{0.5^2 + 1.5^2} = 1.59$$

Let consider 4 sources centered on $(x, y), (x, y-3), (x+1, y), (x+1, y-3)$. If $r = 1.5$ there is an area that is not covered as shown in Figure 10. To cover the space, we must extend the sources coverage radius to 1.59 as show by r′ in Figure 10-B.

**Observarion 5.2.2** Given 4 sources centered on $(x, y), (x, y+3), (x+1, y), (x+1, y+3)$. If all sources have radius equal to 1.59, the area under the rectangular with the top-left-corner-point on $(x, y)$ and the bottom-right-corner-point on $(x+3, y+3)$ will be covered.

**Theorem 5.2.3** *If all sources from ASGP that cover a square grid size of p have radius equal to 1.59, the square area size of p-1 will be covered.*

**Proof:** Because the location of sources for covering a square grid is systematically specified, by considering all the spaces possible according to Observation 5.2.1 and 5.2.2, all the square areas under the square grid will be covered.  □

## 6. CONCLUSIONS AND FUTURE WORKS

In this paper, we propose an approximation algorithm to solve a SQUARE GRID POINTS COVERAGE (*SGPC*) problem which is in the *NP*-complete class. We try to minimize the number of sources with coverage radius of one to cover all points in a square grid with the condition that all sources must be communicable. We know that any number can be written in an addition of 3, 4 and 5, so we present APPROX-SQUARE-GRID-COVERAGE (*ASGC*) algorithm. The algorithm run in $O(p^2)$, where $p$ is the size of the square grid. The algorithm also guarantees the approximation ratio of $1 + \dfrac{2p - 10}{p^2 + 2}$. Moreover, *we* state about the extension usage of our algorithm and show two examples. We prove that if our sources under *ASPC* algorithm for covering a square grid size of $p$ can move, in eight-time-steps movement, the area under the square gird will be covered. We also prove that if we extend our source coverage radius to 1.59, without any movement the area under the square gird will also be covered.

For future research, we conclude our article with a list of open problems.

   – We may vary the radius of sources.
   – We may work on some other types of the grid.
   – We may extend the area into *k*-dimension.
   – We may define the movement steps from a starting position to make the coverage area and guarantee the time and the battery consumption.
   – We may include more objects to our model for more realistic. (Adding some obstacles to block the wireless signal.)
   – We may add movement and velocity to each source to have mobility as in real world.
   – We may improve the algorithm. (better the time bound, tighter with a better analysis)

## ACKNOWLEDGEMENTS

Special thanks to Prof. Raymond Greenlaw and Assoc. Prof. Sanpawat Kantabutra for many useful comments and suggestions. This work receives financial support from the Thailand Research Fund through the Royal Golden Jubilee Ph.D. Program (Grant No. PHD/0003/2551).

## Authors

P. Longani is presently doing Ph.D. in Computer Science Department, Faculty of Science, Chiang Mai University, Thailand, together with being a lecturer in the Collage of Arts, Media and Technology in the same university. Miss Longani get Bechelor of Science (Computer Engineering) degree in 2006 and get Master of Science (Computer Science) degree in 2008 from Chiang Mai University, Thailand. Major research area is complexity theory, algorithm analysis, theory of computation.

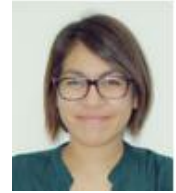